\documentclass{aa}

\usepackage{amsmath}
\usepackage{subfigure}
\usepackage{graphicx}
\usepackage{nicefrac}
\usepackage[dvipsnames]{xcolor}
\usepackage{lipsum}
\usepackage{caption}
\usepackage{dblfloatfix}    
\usepackage{txfonts}
\usepackage[hidelinks]{hyperref}
\usepackage{lipsum}

\hypersetup{colorlinks=true,allcolors=[rgb]{0,0,0.8}}

\usepackage{soul}

\newcommand{\hMpc}{h^{-1}\mathrm{Mpc}}

\newcommand{\hMsun}{h^{-1}\mathrm{M}_\odot}

\newcommand{\vp}{v_\textrm{p}}
\newcommand{\zobs}{z_\textrm{obs}}
\newcommand{\zcos}{z_\textrm{cos}}
\newcommand{\vmax}{V_\textrm{max}}

\begin{document}

   \title{Correlated residuals in Tully–Fisher and Fundamental Plane relations and their impact on peculiar velocity measurements}

   \author{Tyann Dumerchat \inst{1,4}\fnmsep\thanks{E-mail: \href{mailto:tdumerchat@ifae.es}{tdumerchat@ifae.es}}, 
   Raul E. Angulo\inst{2,3}\fnmsep\thanks{E-mail: \href{mailto:reangulo@dipc.org}{reangulo@dipc.org}},
   Julian Bautista \inst{1},
   Cesar Aguayo\inst{4},
    Sownak Bose\inst{5},
    Lars Hernquist\inst{6}
   }

   \institute{
        Aix Marseille Univ, CNRS/IN2P3, CPPM, Marseille, France 
            \and
        Donostia International Physics Center, Manuel Lardizabal Ibilbidea, 4, 20018 Donostia, Gipuzkoa, Spain. 
            \and
        IKERBASQUE, Basque Foundation for Science, 48013, Bilbao, Spain.
            \and
        Universitat Autònoma de Barcelona, IFAE, Bellatera, Spain.
            \and
            LadHyX UMR CNRS 7646, École Polytechnique, Institut Polytechnique de Paris, 91128 Palaiseau Cedex, France.
            \and
        Institute for Computational Cosmology, Department of Physics, Durham University, South Road, Durham DH1 3LE, UK.
        \and
        Harvard-Smithsonian Center for Astrophysics, 60 Garden St, Cambridge, MA 02138, USA.
    }

\date{Received September 15, 1996; accepted March 16, 1997}

   \titlerunning{TF and FP in the MTNG}
   \authorrunning{T. Dumerchat et al.}

  \abstract{
The Tully-Fisher (TF) and Fundamental Plane (FP) relations are widely used to infer extragalactic distances and peculiar velocities, enabling measurements of large-scale velocity statistics and cosmological parameters. Using the Millennium-TNG hydrodynamical simulation, we assess the accuracy of these methods in the presence of realistic galaxy formation physics.
We find that, while the 2-point statistics of velocities are reliably inferred on scales larger than $\sim10\,\hMpc$, significant systematic deviations arise on smaller scales. These deviations originate from spatially correlated residuals in the TF and FP relations, driven by correlations between galaxy structural properties, star-formation history, and the local environment. As a result, TF- and FP-inferred velocity fields exhibit spurious correlations with the galaxy density field that cannot be explained by random scatter alone. We show that extending the TF and FP relations to include additional galaxy properties — such as star formation rate, gas mass, and stellar mass — mitigate these environmental correlations, particularly for late-type galaxies. Our results demonstrate that galaxy formation physics induces significant systematics in peculiar velocity measurements on non-linear scales, and that neglecting these effects may bias cosmological analyses.
}
\maketitle
%

\section{Introduction}
Peculiar velocities in the cosmos arise from gravitational interactions associated with the growth of structure and, therefore, provide a powerful probe of cosmology. Measurements of the peculiar velocity field and its statistical properties constrain the growth rate of structure, test gravity on large scales, and complement other cosmic probes \citep{adamsImprovingConstraintsGrowth2017,turnerLocalMeasurementGrowth2023}. 

Over the past decades, peculiar velocity measurements have relied primarily on empirical distance indicators, most notably the Tully–Fisher \citep[TF,][]{TullyFisher_1977} relation for late-type galaxies and the Fundamental Plane \citep[FP,][]{FundamentalPlane_1987} relation for early-type galaxies. The TF relation correlates the absolute magnitude $M$ of a spiral galaxy  and its maximum rotation velocity. Analogously, the FP relation correlates the effective radius $R$ of an early-type galaxy to its velocity dispersion and surface brightness. Since $M$ and $R$ correspond to intrinsic galaxy properties, they can be respectively compared with the apparent magnitude and size of a galaxy to infer its distance. 

The TF or FP distance estimates can be combined with spectroscopic redshifts to infer the line-of-sight peculiar velocity of galaxies. This approach has been widely used to construct peculiar velocity catalogues \citep{saulder_calibrating_2013, campbell_6df_2014, hong_2mtf_2019, howlett_sloan_2022, tully_cosmicflows-4_2023}. 
The most recent measurements were released by the DESI Peculiar Velocity survey using Data Release 1, achieving a 12\% constraint on the growth rate $f\sigma_8$ through a joint analysis of galaxy and peculiar velocity clustering \citep{bautistaDESIDR1Peculiar2025,carrDESIDR1Peculiar2025,douglassDESIDR1Peculiar2025,laiDESIDR1Peculiar2025,qinDESIDR1Peculiar2025,rossDESIDR1Peculiar2025,turnerDESIDR1Peculiar2025}.

Upcoming data releases, are expected to significantly improve the statistical precision of these measurements \citep{saulderTargetSelectionDESI2023}, which motivates a careful assessment of the systematic errors affecting these methods.
A well-known limitation of the TF and FP relations is their significant intrinsic scatter, which translates into distance uncertainties of approximately 20\%.
In most analyses, this scatter is assumed to be random and uncorrelated. This means that TF- and FP-based velocity estimators are unbiased, i.e. the scatter contributes only random noise to velocity statistics. However, this assumption may not hold in detail. 

Hydrodynamical simulations have shown that galaxy structural and kinematic properties correlate with environment, assembly history, and star formation. This raises the possibility that residuals around the TF and FP relations are not purely stochastic, but instead spatially correlated \citep{Illustris_FP_2020, Illustris_TF_2023,Simba_TF_2020,Eagle_FP_2023,Huang_TF_residual_2024}. Such correlated residuals would induce spurious correlations between inferred peculiar velocities and the galaxy density field, biasing velocity statistics. Despite its potential importance for cosmological analyses, the impact of correlated TF and FP residuals on peculiar velocity measurements has not yet been systematically quantified.

In this work, we investigate the accuracy of TF- and FP-based peculiar velocity estimators using the Millennium-TNG (MTNG) hydrodynamical simulation \citep{MTNG_2023}. MTNG combines a large cosmological volume with sufficient resolution to model the internal properties of galaxies, making it ideally suited for exploring the TF and FP distance estimators. 

By constructing TF and FP relations directly from the simulated galaxies, we will indeed show that correlated residuals do bias velocity statistics on non-linear scales. We further explore the physical origin of correlated TF and FP residuals, and examine the dependence of these relations on other galaxy properties (stellar mass, gas content, star-formation rate, and environment). We then extend the TF and FP relations so that they incorporate additional galaxy parameters, showing that they exhibit reduced scatter and spatial correlations. Although these extended relations are not intended to be directly applicable to observations, they demonstrate the role of galaxy formation physics in shaping TF- and FP-based distance estimates.

Our paper is structured as follows. In Section \ref{sec:mtng}, we describe the MTNG simulation and the samples used in our analysis. In Section \ref{sec:tf_fp}, we construct the TF and FP relations in our mock catalogues. In Section \ref{sec:pv}, we examine the impact of TF- and FP-based distance errors on peculiar velocity statistics. Finally, in Section \ref{mtng:sec:residual}, we explore the physical origin of correlated residuals and connect these findings to galaxy formation histories in Section \ref{mtng:sec:merger}. We present our conclusions in Section \ref{sec:conclusions}.


\section{Mock galaxy catalogues}
\label{sec:mtng}

In this section, we describe our mock galaxy catalogues of late-type and early-type galaxies. These catalogues form the basis for our analysis of the intrinsic scatter of the Tully–Fisher (TF) and Fundamental Plane (FP) relations and their impact on peculiar velocity measurements.

\subsection{The Millennium-TNG simulation}

The Millennium-TNG (MTNG) simulation suite provides a state-of-the-art framework for studying galaxy formation and evolution and their imprint on large-scale structure \citep{MTNG_2023, Hernandez_MTNG_2023, Barrera_MTNG_2023, Kannan_MTNG_2023, Bose_MTNG_2023, Hadzhiyska_MTNG_2023}. MTNG builds upon the Millennium \citep{Millenium_2005} and Illustris-TNG \citep{Illustris_Pillepich_2018, Illustris_Marinacci_2018, Illustris_Springel_2018, Illustris_Nelson_2018, Illustris_Naiman_2018} projects, with the goal of accurately modelling the galaxy–halo connection and the effects of baryonic physics on clustering. 

MTNG was run with the moving-mesh code {\tt AREPO} \citep{Springel_meshcode_2010} and includes a comprehensive treatment of baryonic physics, such as radiative cooling, star formation, supermassive black hole growth, and feedback from supernovae and active galactic nuclei. Dark matter halos and subhalos are identified across 265 snapshots, enabling detailed tracking of merger histories and galaxy evolution. Additionally, the suite includes simulations of varying volumes, along with gravity-only runs employing variance-suppression techniques \citep{anguloCosmologicalBodySimulations2016}. The adopted cosmological parameters in MTNG match those of Illustris-TNG, based on analyses of the Planck satellite and large-scale structure data \citep{Planck_2016}.

In this work, we use the largest hydrodynamical periodic box, with a side length of $500\,\hMpc$. Our analysis focuses on scales below $\sim60\,\hMpc$, which are not expected to be significantly affected by the finite box size. The simulation evolves $4320^3$ dark matter particles and $4320^3$ gas cells, with mass resolutions of $1.7\times10^8\,\hMsun$ and $3.1\times10^7\,\hMsun$, respectively. 

In MTNG, galaxies are defined as stellar particle agglomerations within subhalos. To ensure robust measurements of structural and kinematic properties, we restrict our analysis to subhalos with stellar masses $M_* > 10^9\,\hMsun$ and at least $10^3$ stellar particles. 

\subsection{Peculiar velocities}
\label{mtng:sec:pv} 
We incorporate the effects of peculiar velocities into our mock catalogues as follows. The line-of-sight peculiar velocity $\vp$ shifts the observed redshift $\zobs$ according to

\begin{equation}
1 + \zobs = (1 + \zcos)(1 + \vp/c),
\label{eq:rsd}
\end{equation}

\noindent where $\zcos$ is the cosmological redshift and $c$ is the speed of light. 

Peculiar velocities also affect the inferred absolute magnitude $M$ and physical size of a galaxy $R$ at fixed observed flux and angular size. We therefore compute and store, for each galaxy,

\begin{eqnarray}
M(\zobs) &=& M(\zcos) + 5 \log\frac{D_L(\zobs)}{D_L({\zcos})} \\
\log R(\zobs) &=& \log R(\zcos) + \log\frac{D_A(\zobs)}{D_A({\zcos})}
\end{eqnarray}

\noindent where $D_L$ and $D_A$ are the luminosity and angular diameter distance, respectively. In Appendix~\ref{appendix:TF_FP}, we provide further details on these relations.

\subsection{Sample definition}
\label{sec:data:selection} 

\begin{figure}
    \centering
  \includegraphics[width=1\columnwidth]{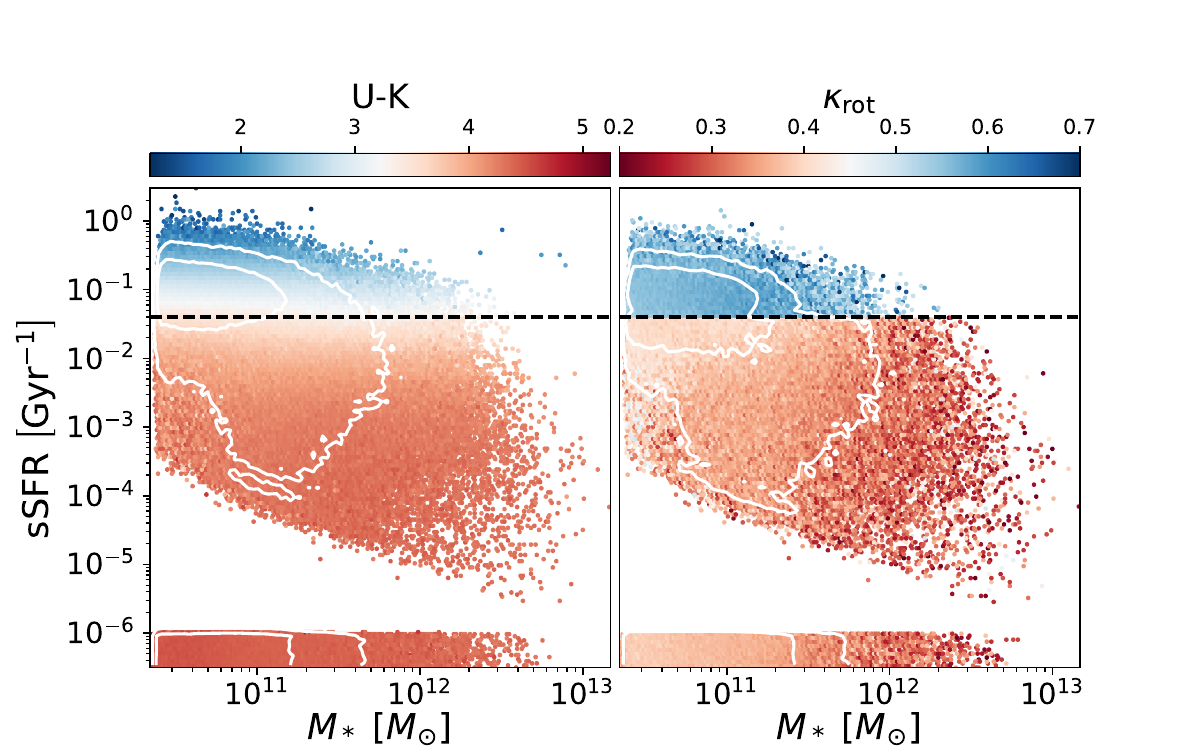} 
  \caption{
  Galaxy sample selection in the MTNG simulation. {\it Left panel}: Specific star formation rate (sSFR) as a function of stellar mass for all galaxies at $z=0$, colour-coded by U–K colour. The black dotted line marks the threshold $\mathrm{sSFR}=0.04\mathrm{Gyr}^{-1}$ used to separate star-forming and quiescent galaxies. White contours enclose the 68\% and 95\% density regions of the distribution.
 {\it Right panel}: Final late-type (LTG) and early-type (ETG) samples after applying an additional cut on the rotational support parameter $\kappa_\mathrm{rot}$. Galaxies are colour-coded by $\kappa_\mathrm{rot}$, highlighting the separation between rotation- and dispersion-dominated systems. }
\label{fig:etg_ltg_selection}
\end{figure}

The TF relation applies to late-type galaxies (LTGs), which are star-forming, blue, and predominantly rotation-supported systems, while the FP relation describes early-type galaxies (ETGs), which are quiescent, red, and primarily supported by velocity dispersion.
  
We distinguish between LTGs and ETGs using their specific star formation rate (sSFR) and the rotational-to-kinetic energy ratio $\kappa_\mathrm{rot}$. The sSFR is defined as $\mathrm{sSFR}=\mathrm{SFR}/M_*$, where the star formation rate is measured in $M_\odot \ \mathrm{Gyr}^{-1}$ and $M_*$ is the stellar mass. The parameter $\kappa_\mathrm{rot}$ is given by:

\begin{equation}
    \kappa_{\text {rot}} =\frac{\sum_n \frac{1}{2} m_n \left(\frac{j_{n}}{r_n}\right)^2 }{ \sum_n \frac{1}{2} m_n \mathbf{|v|}_n^2 },
\end{equation}

\noindent where the sum runs over stellar particles $n$, with mass $m_n$, angular momentum $j_n$ (computed along the galaxy minor axis), distance from the galaxy centre $r_n$, and velocity $\mathbf{v}_n$ in the centre-of-mass frame. Although $\kappa_\mathrm{rot}$ could be computed using projected quantities to better mimic observational procedures, we use three-dimensional quantities to obtain a cleaner characterisation of the intrinsic dynamical state of galaxies.

We classify galaxies as: 

\begin{itemize}
    \item LTGs: sSFR $>0.04 \ \text{Gyr}^{-1}$ and $ \kappa_{\text {rot}} > 0.5$
    \item ETGs: sSFR $<0.04 \ \text{Gyr}^{-1}$ and $ \kappa_{\text {rot}} < 0.5$.
\end{itemize}
\noindent where we set the limit of $0.04$ for sSFR to distinguish galaxies above and below the main sequence of star formation.

These criteria do not exactly match observational definitions, which depend on survey-specific selections and available data. However, they provide a physically motivated separation that captures the main dynamical and star formation properties relevant for TF and FP analyses.

Figure~\ref{fig:etg_ltg_selection} illustrates this selection. The left panel shows the sSFR–stellar mass plane for the full galaxy sample, with the sSFR threshold marked by the black dotted line and colour-coded by U–K, confirming the expected separation between blue, star-forming and red, quiescent systems. The right panel shows the final LTG and ETG samples colour-coded by $\kappa_\mathrm{rot}$, highlighting the distinction between rotation- and dispersion-dominated galaxies.

We select galaxies from the $z=0$ snapshot and construct mock TF and FP catalogues under the flat-sky approximation, shifting galaxy positions along the line of sight to effective redshifts of $z\simeq0.04$ for LTGs and $z\simeq0.08$ for ETGs. These choices are motivated by the expected redshift distributions of ongoing and upcoming TF and FP surveys such as DESI \citep{saulderTargetSelectionDESI2023}. Our final catalogues contain $84,787$ LTGs and $271,358$ ETGs.

Throughout the remainder of this paper, we use blue to present LTGs and TF-related results, and red for ETGs and FP-related results.

\section{TF and FP relations in the MTNG simulation}
\label{sec:tf_fp}

In this section, we construct and characterise the TF and FP relations using our MTNG mock galaxy catalogues. These relations form the basis for our subsequent analysis of residual scatter and its impact on peculiar velocity measurements.

\begin{figure*}
  \centering
  \includegraphics[width=0.8\columnwidth]{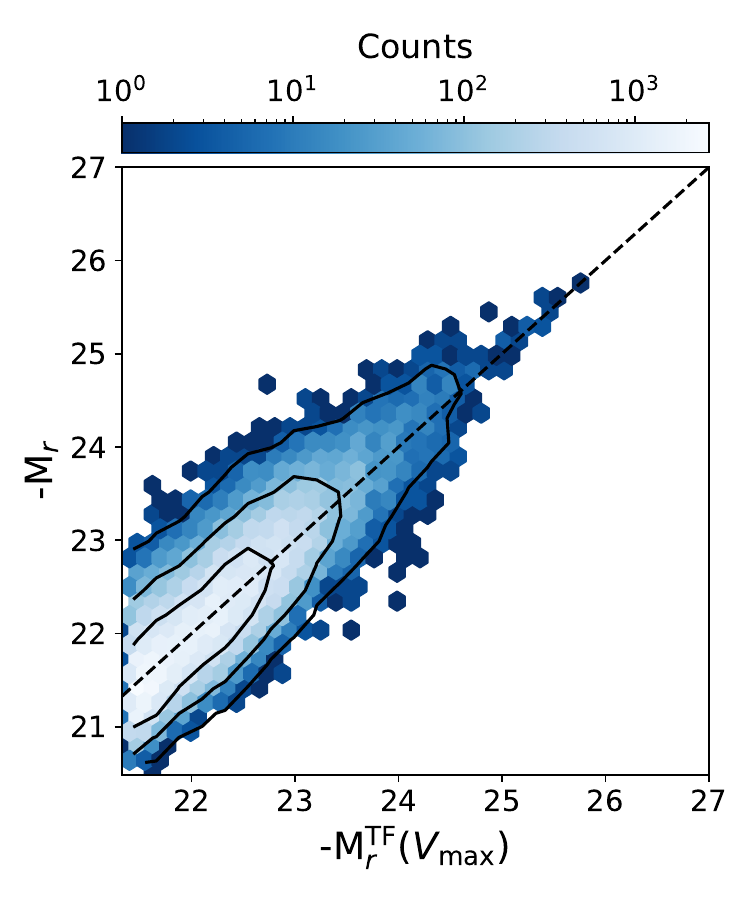} 
  \includegraphics[width=0.8\columnwidth]{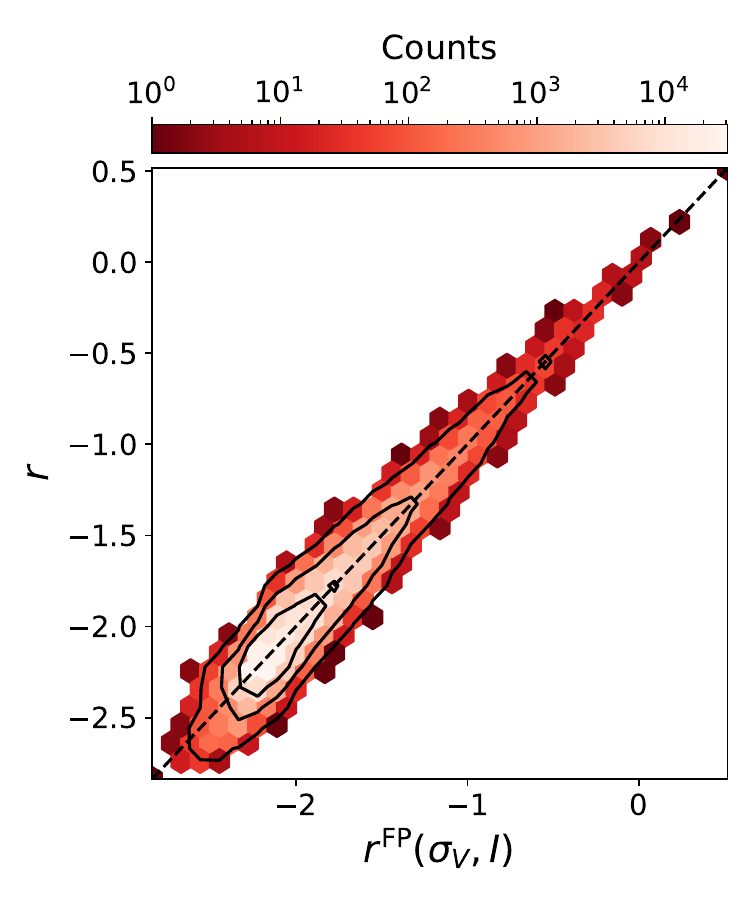} 
\caption{Tully–Fisher and Fundamental Plane relations in the MTNG simulation. Left: Comparison between true $r$-band absolute magnitudes and those predicted by the non-parametric TF relation for late-type galaxies. Right: Comparison between true logarithmic effective radii and values predicted by the non-parametric FP relation for early-type galaxies. In both panels, the black dashed line indicates the one-to-one relation, while contours enclose 68\%, 95\%, and 99.7\% of the galaxy distribution.}
\label{fig:TF_FP_relation}
\end{figure*}

\subsection{Tully-Fisher and Fundamental Plane Relations}

Late-type galaxies (LTGs) follow the Tully–Fisher relation, which links their absolute magnitude $M$ to their maximum rotational velocity $V_\mathrm{max}$. In observational analyses, $V_\mathrm{max}$ is typically inferred from line-of-sight velocities with inclination corrections. In contrast, the simulation allows us to directly measure the maximum circular velocity, defined as $\vmax \equiv \max{\sqrt{G M(<r)/r}}$, where $M(<r)$ is the total mass enclosed within radius $r$. Throughout this work, we adopt the notation $v_\mathrm{max} \equiv \log_{10} V_\mathrm{max}$ and use $r$-band absolute magnitudes $M_r$, assuming perfect dust extinction correction.

Observational TF relations are commonly calibrated using linear regressions on restricted calibration samples. Here, since the simulation provides noise-free measurements of $(M_r, v_\mathrm{max})$ for the full galaxy population, we use the complete LTG sample to suppress selection effects. Furthermore, we relax the assumption of linearity and construct a more flexible relation. Specifically, we define a non-parametric TF relation $M_r^\mathrm{TF}(v_\mathrm{max})$ by interpolating the mean absolute magnitude $\langle M_r \rangle_k$ in bins of $v_\mathrm{max}$. We use this relation to predict the absolute magnitude of any LTG based solely on its rotational velocity.

The Fundamental Plane relation describes early-type galaxies (ETGs) through a correlation between physical effective radius $R$, velocity dispersion $\sigma_V$, and surface brightness $I$. We compute $\sigma_V$ as the three-dimensional velocity dispersion within $R$ and approximate the effective radius using the stellar half-mass radius. The surface brightness is defined as

\begin{equation}
    I = \frac{L}{\pi R^2},
\end{equation}

\noindent where $L$ is the galaxy luminosity. We adopt the standard convention and use the variables $r \equiv \log R$, $s \equiv \log \sigma_V$, and $i \equiv \log I$. 

Following the same strategy as for the TF relation, we construct a non-parametric FP relation by interpolating the mean effective radius $\langle r \rangle_k$ in bins of $(s,i)$. This allows us to predict the physical size of ETGs from their velocity dispersion and surface brightness without assuming a specific functional form.

\subsection{Performance of non-Parametric TF and FP relations}

Figure~\ref{fig:TF_FP_relation} illustrates the performance of the non-parametric TF and FP relations. The left panel compares the true $r$-band magnitudes, $M_r$, with those predicted by $M_r^\mathrm{TF}(v_\mathrm{max})$, while the right panel shows the true logarithmic radii, $r$ against the FP predictions $r^\mathrm{FP}(s,i)$. In both cases, the relations are unbiased by construction but exhibit significant intrinsic scatter arising from galaxy formation physics.

To quantify the impact of this scatter on distance measurements, we define the log-distance ratios:

\begin{eqnarray}
    \eta_L &\equiv& \log \frac{D_L(\zobs)}{D_L(\zcos)} =0.2 \left[M_r(\zcos) - M_r^\textrm{TF}(\zobs) \right] \\
    \eta_A &\equiv& \log \frac{D_A(\zobs)}{D_A(\zcos)} =r^\textrm{FP}(\zobs) - r(\zcos) .
\end{eqnarray}

\noindent where $D_L$ and $D_A$ are the luminosity and angular diameter distances, respectively.

Since $M_r^\mathrm{TF}$ and $r^\mathrm{FP}$ are imperfect estimators of the true galaxy properties, the observed log-distance ratios can be written as

\begin{equation}
  \eta^\textrm{obs}_L  = \eta_L  + \lambda_\textrm{TF}, \quad \text{and} \quad
  \eta^\textrm{obs}_A  = \eta_A  + \lambda_\textrm{FP},
\label{eq:eta_residual}
\end{equation}
\noindent where $\lambda_\mathrm{TF}$ and $\lambda_\mathrm{FP}$ denote the residuals of the TF and FP relations, respectively.

For the TF sample, we find $\mathrm{std}(\lambda_\mathrm{TF}) = 0.062$, corresponding to a fractional distance error of approximately $14\%$. Recent observational TF catalogues report scatters of order $22\%$ \citep{hong_2mtf_2019,tully_cosmicflows-4_2023}. For the FP sample, we measure $\mathrm{std}(\lambda_\mathrm{FP}) = 0.075$, corresponding to a fractional distance error of $\sim17\%$, compared to $\sim23\%$ reported for SDSS-based FP catalogues \citep{saulder_calibrating_2013,howlett_sloan_2022}.

The agreement between simulated and observed scatter is remarkable, especially given that observational measurements are additionally affected by dust attenuation, inclination corrections, and limited calibration samples. This demonstrates that the MTNG mock catalogues provide a realistic and controlled environment to explore the origin and impact of TF and FP residuals on peculiar velocity estimates.

\section{Peculiar velocity clustering}
\label{sec:pv}

In this section, we investigate how astrophysical processes and the intrinsic scatter in the TF and FP relations impact the statistical properties of inferred peculiar velocities. In particular, we assess whether deviations from the idealised assumption of purely random scatter introduce systematic effects in velocity clustering statistics.

\subsection{Inferred velocities and velocity residuals}

We estimate peculiar velocities, $\vp^\mathrm{obs}$, from the observed logarithmic distance ratios $\eta^\mathrm{obs}$ (see Eq. \ref{eq:eta_TF_FP}). We define the corresponding velocity residual as

\begin{equation}
\lambda_v \equiv \vp^\textrm{ obs} - \vp \approx
\begin{cases} 
  \lambda_\textrm{ TF} / \alpha_L(z_\textrm{obs}) & \text{for LTG }  \\
  \lambda_\textrm{ FP} / \alpha_A(z_\textrm{obs}) & \text{for ETG } 
\end{cases}
\label{eq:vel_residuals}
\end{equation}

\begin{figure}
  \centering
\includegraphics[width=0.8\columnwidth]{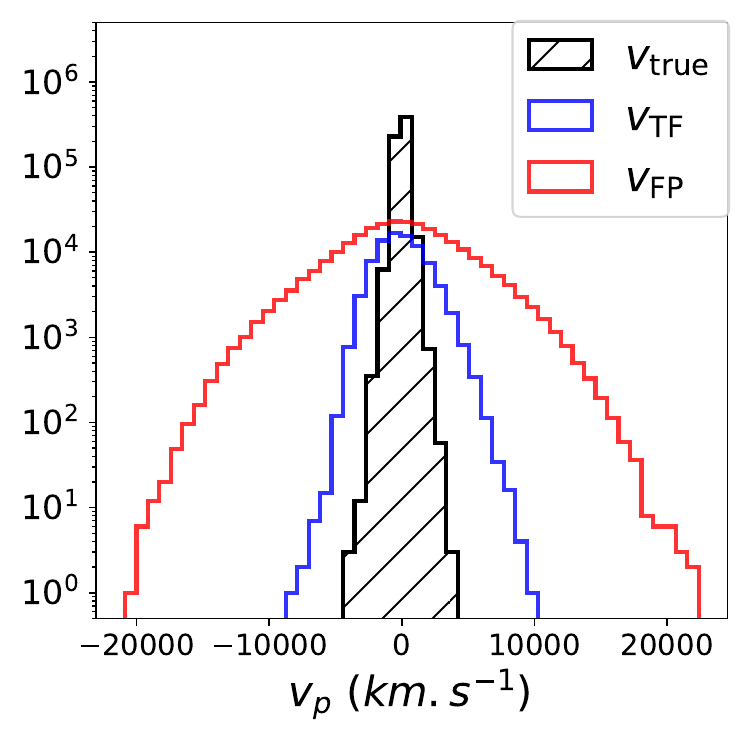} 
\caption{Radial velocity distributions of the galaxy samples. Blue and red histograms show the inferred line-of-sight peculiar velocities for the LTG (TF) and ETG (FP) samples, respectively. The true velocity distribution of the combined samples measured directly from the simulation is shown in black. The broadening of the inferred distributions reflects the dominant contribution of distance–indicator scatter.}
\label{fig:noisy_vel_hist}
\end{figure}

Figure \ref{fig:noisy_vel_hist} shows the distribution of inferred velocities for the TF (blue) and FP (red) samples, compared to the true velocity distribution (black). In both cases, the signal-to-noise ratio is poor, i.e. the velocity uncertainties are much larger than true amplitude of the velocity. While the intrinsic velocity distributions of the two samples are expected to differ, the broadening of the inferred distributions is primarily driven by the scatter in the FP and TF relations.

The noise is more significant for the FP sample for two reasons. First, the intrinsic scatter of the FP relation is larger than that of the TF relation, $\lambda_\mathrm{FP} > \lambda_\mathrm{TF}$. Second, and more importantly, the FP sample extends to higher redshift by construction, which increases the velocity uncertainty for a fixed scatter in distance. In the next subsection, we will explore how these uncertainties affect the velocity clustering.

\begin{figure*}
  \centering
\includegraphics[width=1.5\columnwidth]{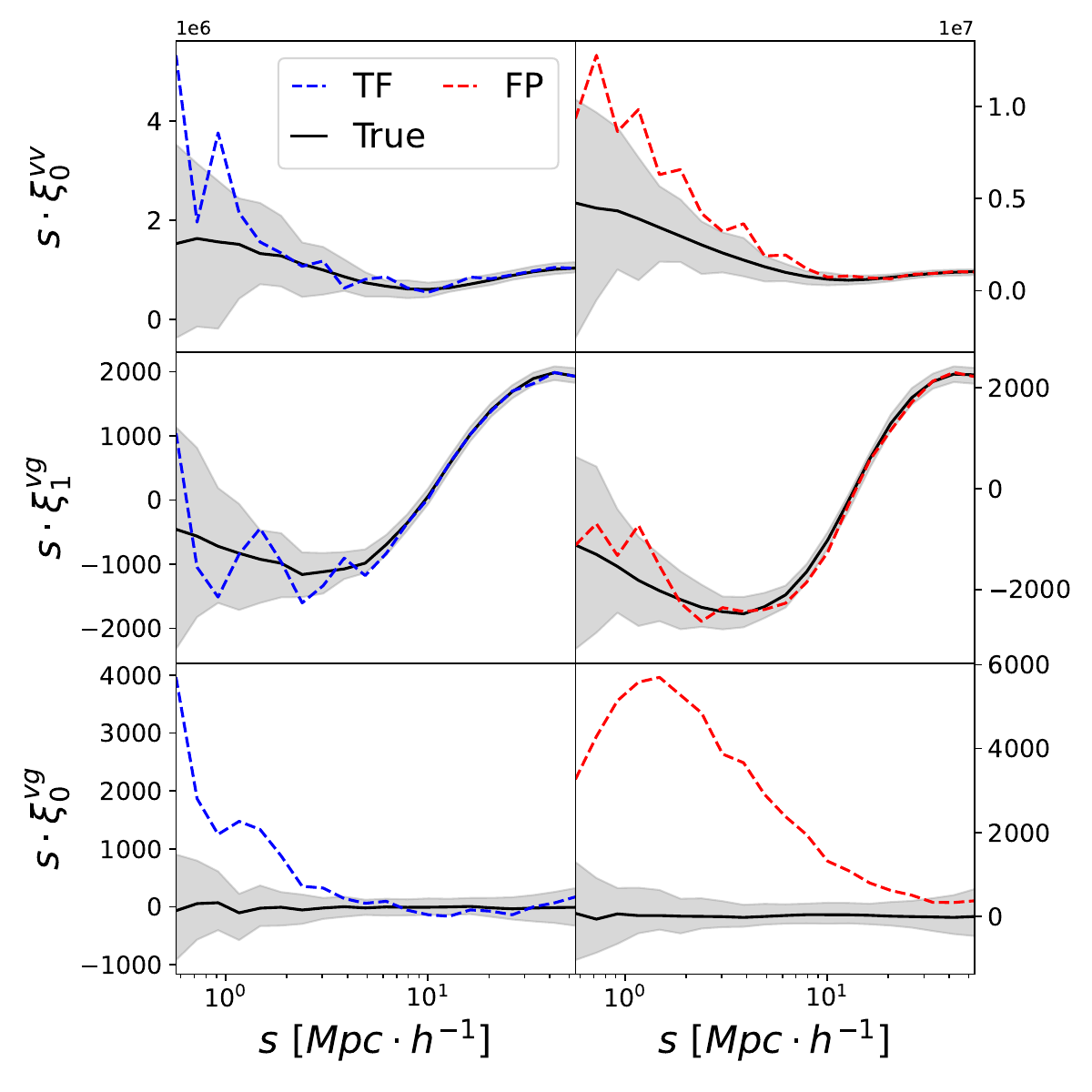} 
\caption{Velocity clustering statistics inferred from TF and FP distances. Blue and red curves show measurements obtained using TF (LTG) and FP (ETG) samples, respectively, while black curves correspond to clustering measured using the true velocities from the simulation. From top to bottom, the panels display the monopole of the velocity auto-correlation $\xi^{vv}_0$, the dipole of the velocity–galaxy cross-correlation $\xi^{vg}_1$, and the monopole of the velocity–galaxy cross-correlation $\xi^{vg}_0$. Measurements are averaged over three orthogonal lines of sight obtained by rotating the simulation box. The shaded regions indicate the $1\sigma$ scatter expected from purely random distance errors, estimated from 30 realisations per line of sight by adding Gaussian noise to the true velocities with standard deviations std$(\lambda_{\mathrm{TF}}) = 0.062$ (left) and std$(\lambda_{\mathrm{FP}}) = 0.075$ (right).}
\label{fig:noisy_slustering}
\end{figure*}
\subsection{Velocity clustering statistics}

We measure the clustering of velocities, $\vp$, and galaxy overdensities, $\delta_g$, in our the LTG and ETG mock samples using the velocity auto-correlation function, $\xi^{vv} \equiv \langle v_p\,v_p\rangle$, and the velocity–galaxy cross-correlation function, $\xi^{vg} \equiv \langle v_p\,\delta_g\rangle$, in redshift space (see appendix \ref{appendix:estimator}).
We decompose these correlation functions in terms of their multipoles:

\begin{equation}
\xi^{i,j}_\ell(s) \equiv \frac{2\ell+1}{2}\int {\rm d}\mu\,\xi^{i,j}(s,\mu)\,P_\ell(\mu),
\label{eq:legendre2}
\end{equation}

\noindent where $\xi^{ij}(s,\mu)$ is the correlation function of two fields $i$ and $j$, $s$ is the pair separation, $\mu$ is the cosine of the angle between the separation vector and the observer’s LoS, and $P_\ell(\mu)$ are Legendre polynomials. To reduce cosmic variance, our measurements are averaged over three orthogonal LoS obtained by rotating the simulation box \citep[e.g.][]{Smith_boxlos_2021}.

Figure \ref{fig:noisy_slustering} presents the clustering measurements using either $\vp^\mathrm{obs}$ inferred from the TF (left, blue) or FP (right, red) relations, or the true $\vp$ measured in MTNG (black). The top, middle, and bottom panels show the monopole of the velocity auto-correlation, $\xi^{vv}_0$, the dipole of the velocity–galaxy cross-correlation, $\xi^{vg}_1$, and the monopole of the velocity–galaxy cross-correlation, $\xi^{vg}_0$, respectively. The shaded regions indicate the standard deviation of clustering measurements obtained using 30 catalogues where Gaussian random noise was added to the true velocities, with a scatter consistent with the uncertainty measured for our TF and FP relations. 

For $\xi^{vv}_0$, both the TF and FP samples recover the expected clustering signal on large scales, where the scatter averages out. On smaller scales, however, significant deviations appear: for the TF sample at $s\sim 1\,\hMpc$ and for the FP sample at $s \sim 10\,\hMpc$, exceeding the $1\sigma$ expectation from purely random noise. In contrast, the dipole $\xi^{vg}_1$ remains consistent with the random-noise expectation at all scales, barely reaching the $1\sigma$ level even on the smallest separations.

The most striking result is found for the monopole of the velocity-galaxy correlation function, $\xi^{vg}_0$. While the true velocities yield $\xi^{vg}_0 \simeq 0$, as expected since these quqantities should be locally uncorrelated, both the TF and FP samples exhibit a highly significant non-zero correlation, exceeding $3\sigma$ for the TF sample and $5\sigma$ for the FP sample.

\subsection{FP and TF correlated residuals}

Using Eq. \ref{eq:vel_residuals}, the observed two-point functions can be written as
\begin{equation}
\begin{aligned}
\xi^{v^\mathrm{obs} v^\mathrm{obs}} &= \xi^{vv} + \xi^{\lambda_v \lambda_v} + \xi^{v \lambda_v} + \xi^{\lambda_v v}, \\
\xi^{v^\mathrm{obs} g} &= \xi^{vg} + \xi^{\lambda_v g},
\end{aligned}
\end{equation}
\noindent where spatial dependencies have been omitted for clarity.

The behaviour of $\xi^{vg}_1$ indicates that the dipole of the residual–galaxy correlation vanishes, $\xi^{\lambda_v g}_1 \approx 0$. The correlated residuals, therefore, affect only the monopole. Additionally, since $\xi^{vg}_0 \approx 0$ for the true velocities, the observed non-zero signal directly implies $\xi^{\lambda_v g}_0 \neq 0$. This implies that the residuals in the TF and FP relations cannot be described as purely random Gaussian noise, but instead correlate with the local density field.

Interestingly, $\xi^{\lambda_v g}_0$ can in principle be measured directly from observational data, since $\xi^{v^\mathrm{obs} g}_0 \approx \xi^{\lambda_v g}_0$. Although this multipole is not used in standard cosmological analyses, it provides a convenient diagnostic for detecting and quantifying correlated residuals. As a proof-of-concept, we have measured this statistics in the SDSS survey, which we present and briefly discuss in Appendix \ref{appendix:SDSS}. We leave a detailed exploration of this approach to future work. 

In the velocity auto-correlation monopole, $\xi^{v^\mathrm{obs} v^\mathrm{obs}}_0$, we also detect non-zero contributions of correlated noise on small scales (below $\sim1\,\hMpc$ for TF and $\sim10\,\hMpc$ for FP). While not shown here, we have verified that $\xi^{v \lambda_v}_0 \approx 0$, indicating that the signal is entirely driven by the residual auto-correlation $\xi^{\lambda_v \lambda_v}_0 \neq 0$.

Overall, our findings are consistent with a residual noise field that traces the underlying matter or galaxy density field. This field could be described perturbatively, as proposed by \citet{Joachimi_FP_residual_2015}, and potentially their associated bias parameters could be calibrated observationally or with numerical simulations. This approach could help quantify and mitigate the effect of correlated residuals, which could be particularly important for exploiting velocity data on nonlinear scales. 



\section{Dependence of residuals on environment}
\label{mtng:sec:residual}

\begin{figure}
  \centering
\includegraphics[width=0.9\columnwidth]{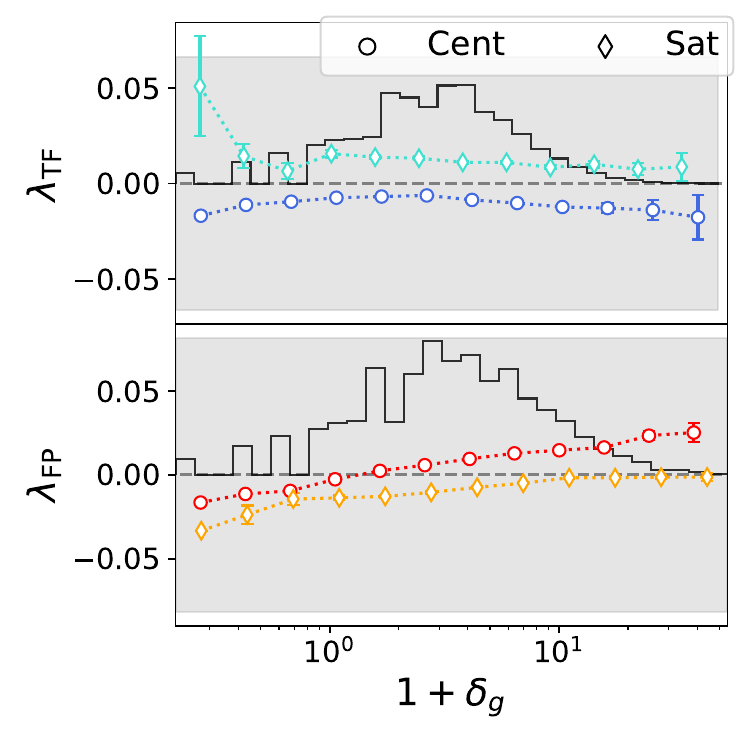} 
\caption{Environmental dependence of TF and FP residuals. Mean residuals in log-distance ratio are shown as a function of the galaxy overdensity $1+\delta_g$. The top and bottom panels correspond to TF ($\lambda_\mathrm{TF}$) and FP ($\lambda_\mathrm{FP}$) residuals, respectively. Circles and diamonds denote central and satellite galaxies. Error bars indicate the standard error on the mean in each bin. Shaded bands show the total scatter of the residuals, std$(\lambda_\mathrm{TF}) = 0.062$ and std$(\lambda_\mathrm{FP}) = 0.075$. Histograms indicate the overdensity distributions of the corresponding samples. }
\label{fig:residual_env}
\end{figure}

The physical motivation of the TF and FP relations is rooted in gravitational dynamics. For LTGs, the TF relation reflects the balance between gravity and rotational support, while for ETGs, the FP relation encodes the effects of virial equilibrium and merger-driven evolution. Both relations can be understood as consequences of the virial theorem, with galaxy mass or size predicted from a small set of structural parameters. Deviations from the simplest virial expectations arise from processes such as energy dissipation during mergers, mass loss, and feedback, and are absorbed into the free parameters used to calibrate the relations.

A natural hypothesis to explain the correlated scatter observed in the TF and FP relations in Section \ref{sec:pv} is that present-day galaxy properties -- such as mass profiles, kinematics, and stellar populations -- reflect their formation histories and are therefore correlated with environment. In an idealised scenario where all relevant structural information is known, luminosity-related quantities such as absolute magnitude or effective radius should be predictable up to stochastic astrophysical fluctuations. In practice, standard TF and FP relations rely primarily on $V_\mathrm{max}$ or $\sigma_V$ (and surface brightness), which do not fully capture the complexity of galaxy structure. As a result, residuals around these relations may retain correlations with the local matter density field. We test this hypothesis by analysing the dependence of TF and FP residuals on environment and other galaxy properties.

\subsection{Dependence on galaxy properties and environment}

\begin{figure*}
  \centering
  \makebox[\textwidth][c]{
\includegraphics[width=0.49\columnwidth]{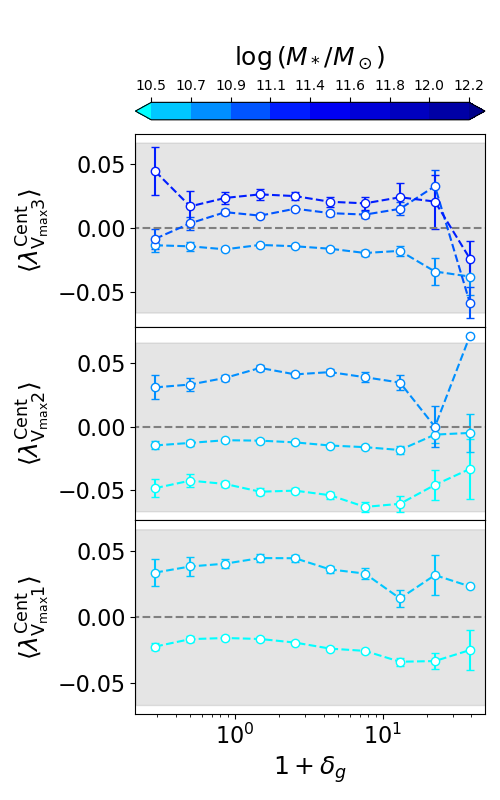} 
\includegraphics[width=0.49\columnwidth]{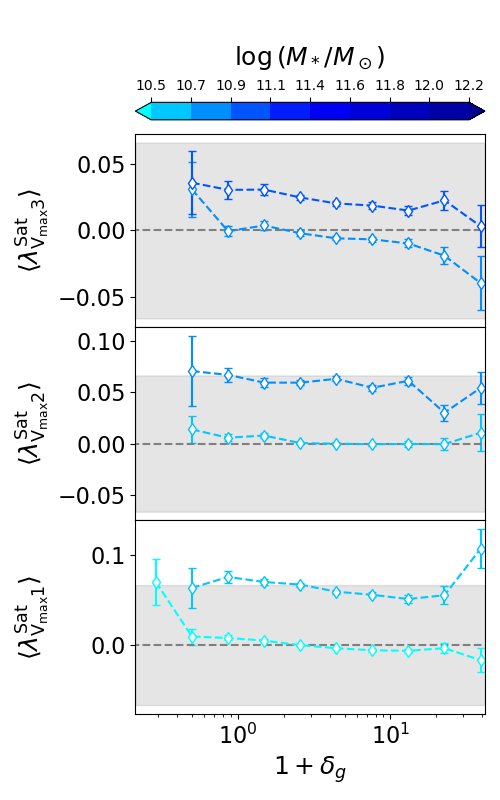} 
\includegraphics[width=0.49\columnwidth]{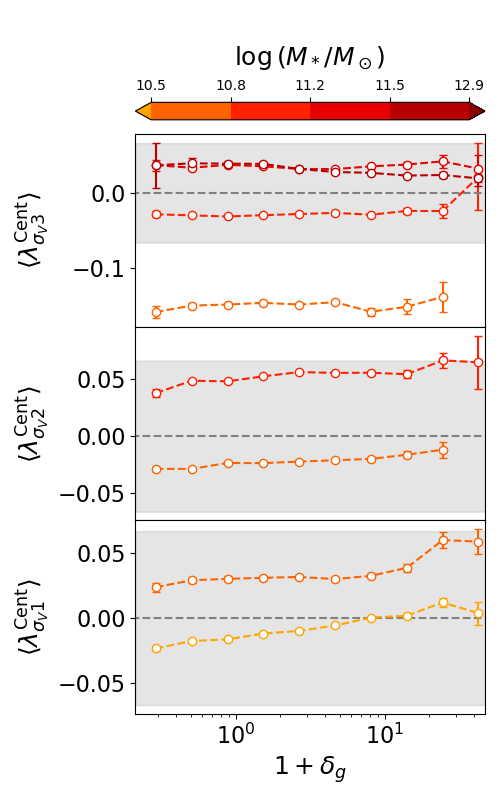} 
\includegraphics[width=0.49\columnwidth]{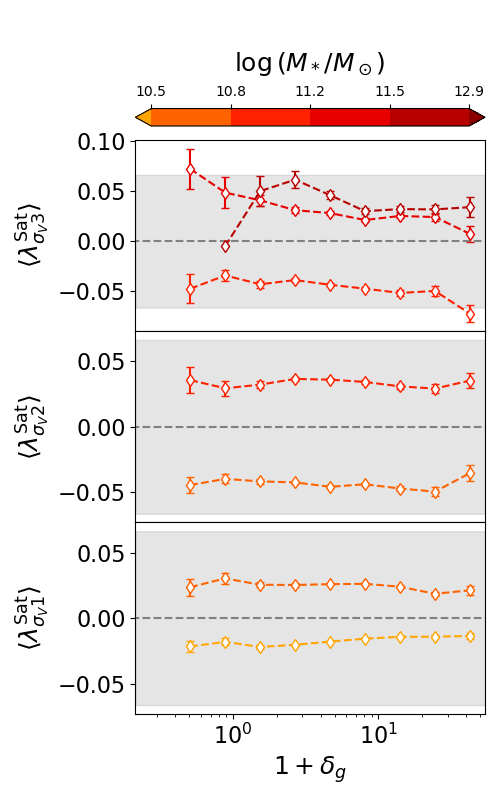} }
\makebox[\textwidth][c]{
\includegraphics[width=0.49\columnwidth]{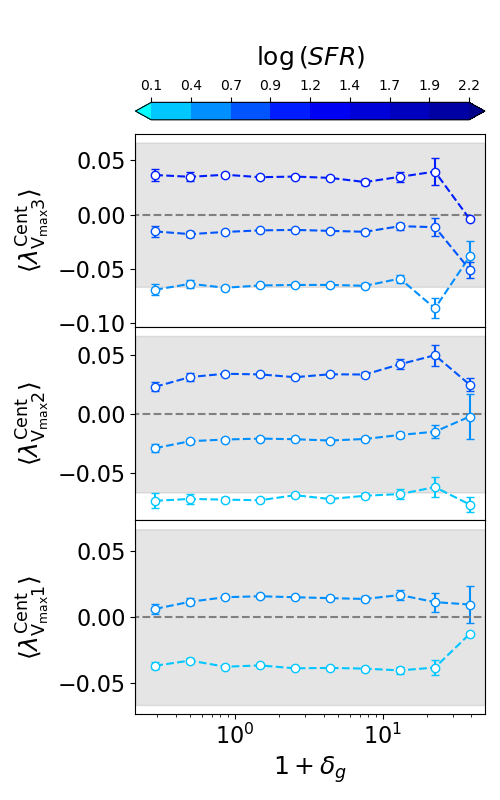} 
\includegraphics[width=0.49\columnwidth]{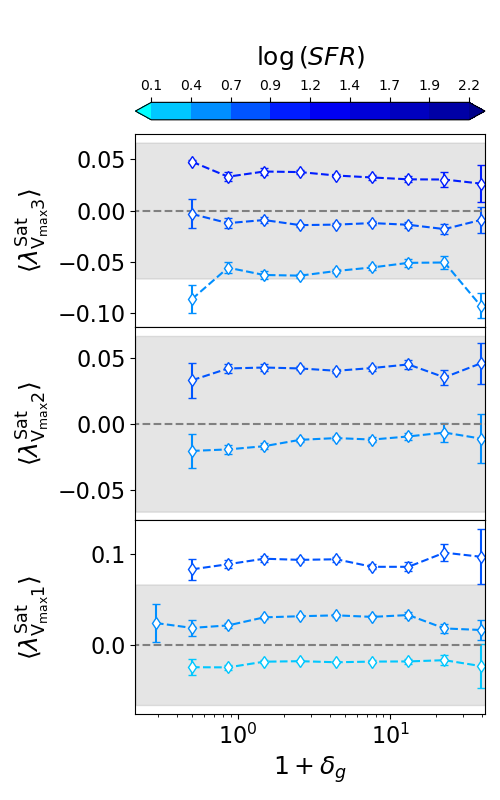} 
\includegraphics[width=0.49\columnwidth]{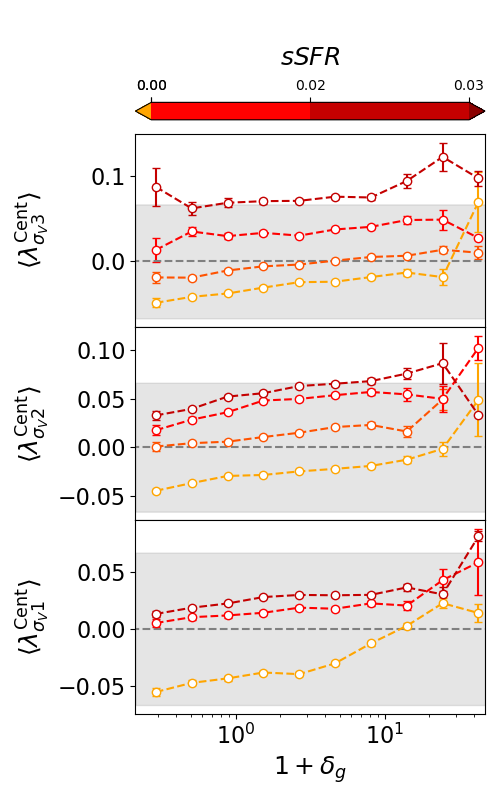} 
\includegraphics[width=0.49\columnwidth]{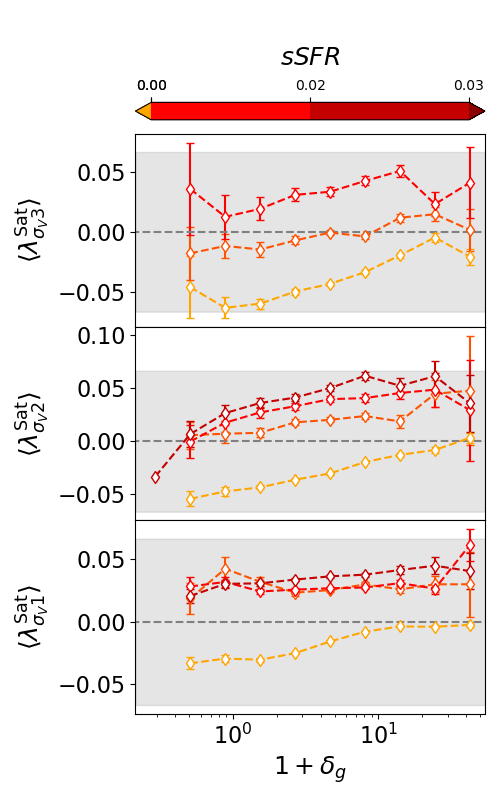}}
\caption{Dependence of TF and FP residuals on galaxy properties and environment. Mean residuals in log-distance ratio are shown as a function of overdensity for LTGs (TF; blue) and ETGs (FP; red), split into bins of $V_\mathrm{max}$ (TF) or $\sigma_V$ (FP). Results are shown separately for central (circles) and satellite (diamonds) galaxies. In each sub-panel, the stacked plots correspond to increasing bins of $V_\mathrm{max}$ or $\sigma_V$ from bottom to top. Shaded bands indicate the total residual scatter, std$(\lambda_\mathrm{TF}) = 0.062$ and std$(\lambda_\mathrm{FP}) = 0.075$.}
\label{fig:residual_properties}
\end{figure*}

We use the full galaxy sample — prior to applying the sSFR and $\kappa_\mathrm{rot}$ selections — to define an environmental indicator based on the galaxy overdensity. Galaxies are assigned to a $256^3$ grid, the galaxy number density $n_g(\boldsymbol{x})$ is smoothed with a spherical top-hat kernel of characteristic scale $\sim 5\,\hMpc$, and the overdensity is computed as $\delta_g(\boldsymbol{x}) = n_g(\boldsymbol{x})/\bar{n} - 1$. Our results are insensitive to the exact choice of smoothing scale in the range $2$–$15\,\hMpc$. We further split the TF and FP samples into central and satellite galaxies, as these populations inhabit different environments and experience distinct evolutionary pathways.

Figure \ref{fig:residual_env} shows the mean TF and FP residuals, $\lambda$, as a function of $1+\delta_g$. Circles and diamonds denote central and satellite galaxies, respectively, with error bars corresponding to the standard error on the mean. The shaded bands indicate the total scatter of the residuals, std$(\lambda_\mathrm{TF}) = 0.062$ and std$(\lambda_\mathrm{FP}) = 0.075$, while the histograms display the overdensity distributions of the samples.

For both the TF and FP relations, central and satellite galaxies occupy systematically distinct regions of the residual space. For LTGs, satellite galaxies preferentially lie above the TF relation, yielding positive residuals and leading to an underestimation of their intrinsic brightness and inferred peculiar velocities (with $\lambda_\mathrm{TF} = 0.2,(M_r^{\mathrm{TF}} - M_r^{\mathrm{true}})$). Central LTGs exhibit the opposite behaviour, with brightness and velocity being overestimated.

For ETGs, satellite galaxies generally show negative FP residuals and lie below the Fundamental Plane, resulting in an overestimation of their effective radii and inferred velocities (with $\lambda_\mathrm{FP} = r_{\mathrm{true}} - r_{\mathrm{FP}}$). In the highest-density environments, the mean satellite residual approaches zero. Central ETGs display a complementary trend, with radii overestimated in underdense regions and underestimated in overdense regions.

Separating central and satellite populations accounts for only a small fraction of the total scatter and does not remove the correlation with environment. In particular, FP residuals for ETGs exhibit a strong dependence on overdensity, increasing systematically with $1+\delta_g$, whereas the TF residuals for LTGs show little to no environmental correlation.

To further investigate the origin of these trends, we examine the dependence of residuals on additional galaxy properties. For the TF relation, it is well known that replacing $V_\mathrm{max}$ with the total baryonic mass can reduce the intrinsic scatter (the baryonic TF relation; \citealt{BTF_2000}). We therefore split the TF central and satellite samples into three bins of $V_\mathrm{max}$ with equal percentiles and study the residual–density correlation within bins of stellar mass or SFR. An analogous procedure is applied to the FP samples, which are divided into bins of $\sigma_V$ and analysed as a function of stellar mass and sSFR.

\begin{figure*}
  \centering
\includegraphics[width=0.9\columnwidth]{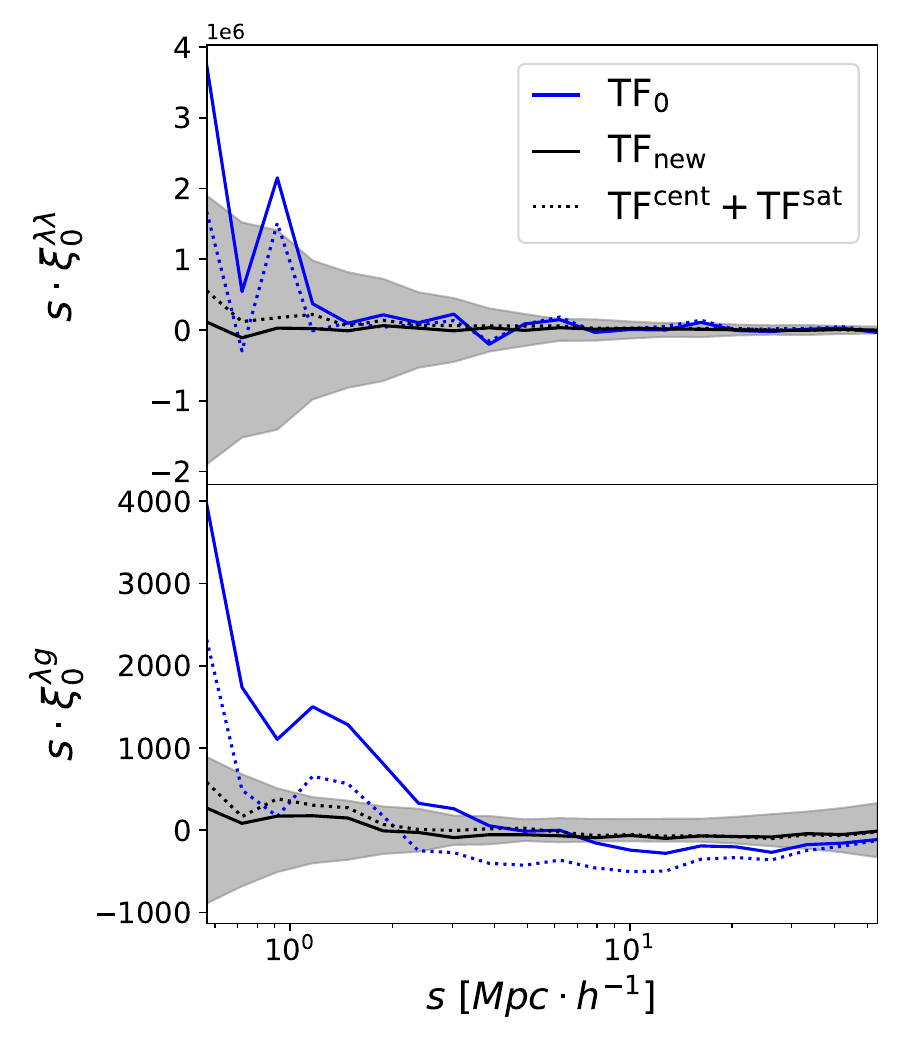} 
\includegraphics[width=0.9\columnwidth]{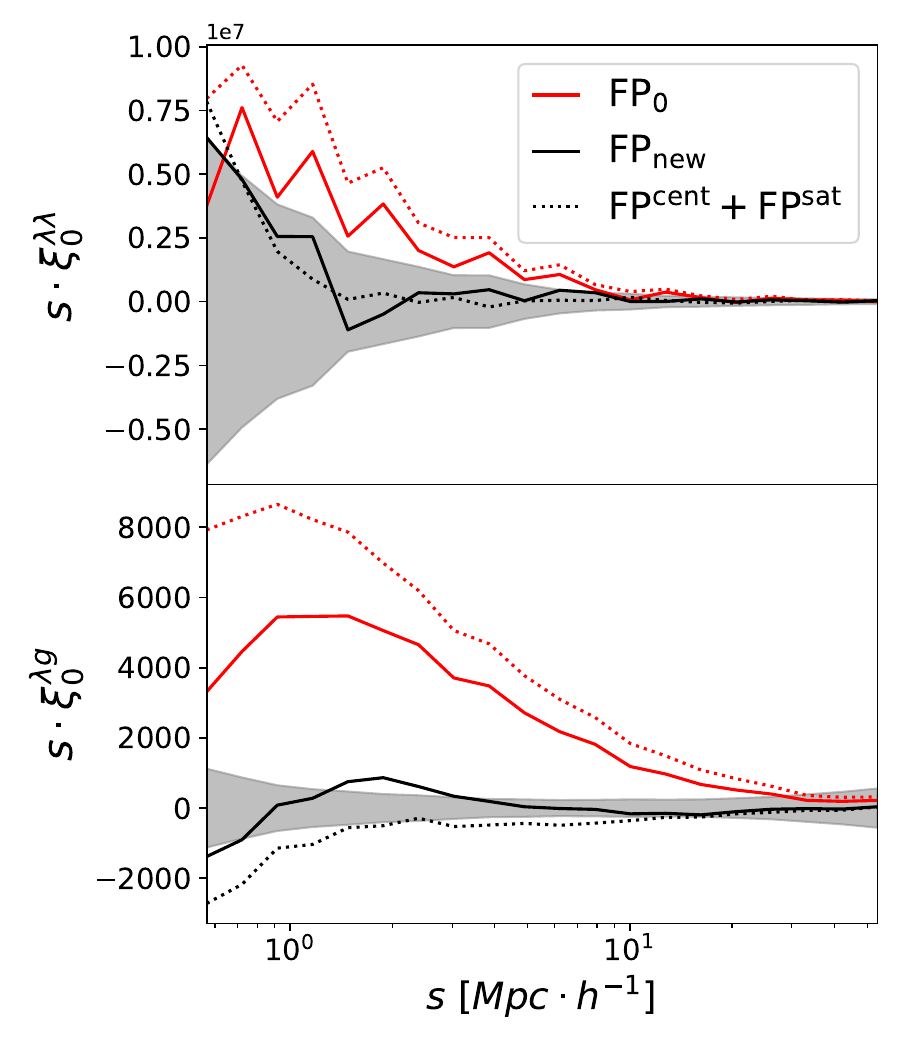} 
\caption{Clustering of TF and FP residuals for standard and extended relations. Two-point correlation functions of the residuals $\lambda$ in log-distance ratio are shown for LTGs (left) and ETGs (right), comparing the standard TF/FP relations (“0”) with the extended models (“new”). Solid lines correspond to relations calibrated on the full samples, while dotted lines show results obtained by calibrating centrals and satellites separately. Grey shaded regions indicate the $1\sigma$ scatter expected from purely random Gaussian residuals, estimated from 30 realisations per line of sight using std$(\lambda_\mathrm{TF}) = 0.062$ and std$(\lambda_\mathrm{FP}) = 0.075$. Clustering measurements are averaged over three orthogonal lines of sight.}
\label{fig:clust_extended_TFFP}
\end{figure*}

Figure \ref{fig:residual_properties} presents the results for LTGs (blue) and ETGs (red), following the same conventions as before. In each sub-panel, the three stacked plots correspond to increasing bins of $V_\mathrm{max}$ (TF) or $\sigma_V$ (FP).

For the TF relation, the dominant contribution to the residual amplitude arises from unaccounted dependencies on stellar mass and SFR. At fixed $V_\mathrm{max}$, galaxies with larger (smaller) stellar mass or SFR have their magnitudes underestimated (overestimated). Once grouped by stellar mass or SFR, the residual distributions of central and satellite galaxies become more consistent with each other, and the dependence on environment is significantly reduced. This indicates that much of the correlated scatter originates from variations in stellar properties not captured by the standard TF relation.

The remaining weak correlations with overdensity may reflect residual trends of stellar mass and SFR within the bins considered. This suggests that, even at fixed $V_\mathrm{max}$, the stellar content of LTGs - and hence their luminosity - retains some dependence on small-scale density fluctuations. A generalised TF relation of the form $M_r^\mathrm{TF}(V_\mathrm{max}, M_*, \mathrm{SFR})$ should therefore reduce the correlated residuals identified in Figure \ref{fig:noisy_slustering}.

We find qualitatively similar behaviour for the FP relation. To mitigate the strong skewness of the SFR distribution due to quenched galaxies, we analyse FP residuals as a function of sSFR. As shown in the right panels of Fig. \ref{fig:residual_properties}, the residual amplitude and the offset between centrals and satellites are largely explained by variations in stellar mass and sSFR within bins of $\sigma_V$. Nevertheless, some dependence on environment remains in specific bins, particularly when grouping by sSFR. This may simply reflect the relatively large width of the sSFR bins and residual correlations between internal properties and environment at fixed $\sigma_V$.

\subsection{Extended TF and FP relations}

Motivated by these findings, we explore extended TF and FP relations that incorporate additional galaxy properties in order to better capture the structural information relevant for predicting luminosity and effective radius. Using the full LTG and ETG samples, we calibrate non-parametric models by interpolating the mean magnitude or radius in multidimensional bins. Specifically, we construct: $M_r^\textrm{ TF}\left(V_\textrm{ max},M_*,M_\textrm{ gas},\textrm{SFR}\right)$ and $r^\textrm{ FP}\left(\sigma_V,I,M_*,M_\textrm{ gas},\textrm{sSFR}\right)$, using logarithmic variables for all quantities except sSFR. 

We emphasise that, for observational applications, TF and FP relations should be calibrated using distance-independent quantities to avoid systematic biases. The extended relations considered here are therefore not intended for direct use on real data, but rather serve as a proof of principle to assess how much of the correlated scatter can be attributed to missing structural information.

Figure \ref{fig:clust_extended_TFFP} compares the clustering of the residuals obtained using the standard and extended TF and FP relations. Results for LTGs and ETGs are shown in the left and right panels, respectively. Dotted lines indicate clustering measurements obtained by calibrating the relations separately for central and satellite galaxies. As before, shaded regions show the $1\sigma$ scatter expected from random Gaussian residuals with std$(\lambda_\mathrm{TF}) = 0.062$ and std$(\lambda_\mathrm{FP}) = 0.075$, and clustering is averaged over three orthogonal lines of sight. 

\begin{figure*}
  \centering
\includegraphics[width=1.7\columnwidth]{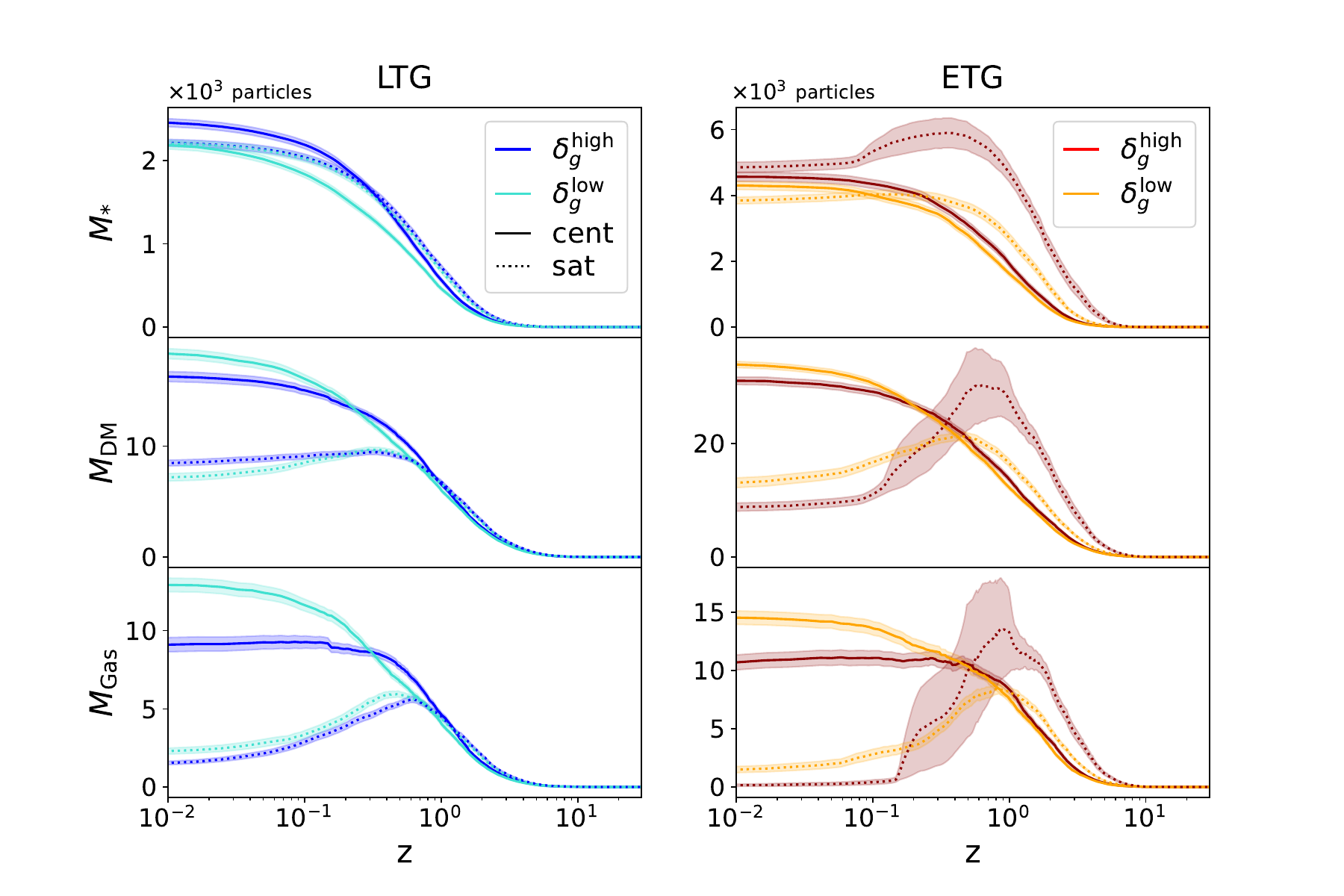} 
\caption{ Redshift evolution of mass components for galaxies in different environments. Median evolution of stellar, dark matter, and gas mass (expressed as the number of simulation particles) for LTGs (left) and ETGs (right), selected within narrow ranges of $v_\mathrm{max}$ and $\sigma_V$, respectively. Solid and dotted lines show central and satellite galaxies, while darker (lighter) colours correspond to galaxies residing in high- (low-) density environments. Shaded regions indicate the standard deviation across 100 galaxies per subsample. From top to bottom, panels show the evolution of stellar, dark matter, and gas mass.}
\label{fig:merger}
\end{figure*}

The extended TF and FP relations substantially reduce the residual scatter and suppress the correlated clustering signal, particularly for the TF case. For the TF relation, splitting centrals and satellites slightly reduces small-scale correlations in the standard model, but has little impact once the extended model is adopted. For the FP relation, separating centrals and satellites unexpectedly enhances the residual clustering for both the standard and extended models, despite reducing the overall scatter. This suggests that, although central and satellite ETGs follow different evolutionary pathways, their structural differences do not trace cosmological density fluctuations on the scales considered here.





To further test this interpretation, we now examine how environment influences the evolutionary histories of the different galaxy populations considered above.

\section{Star formation history}
\label{mtng:sec:merger}

The MTNG simulation provides merger trees, allowing us to trace the assembly histories of individual galaxies. In this section, we investigate how the evolution of gas, dark matter, and stellar mass of a subset of example galaxies depends on large-scale environment, and how these trends help explain the correlated residuals identified in Sections \ref{sec:pv} and \ref{mtng:sec:residual}.

To isolate environmental effects beyond the primary variables entering the TF and FP relations, we select galaxies within narrow ranges of their defining kinematic quantities. Specifically, we retain LTGs with $v_\mathrm{max} \in [2.35,2.38]$ and ETGs with $\log_{10}(\sigma_V) \in [2.12,2.19]$, corresponding to the 40th–60th percentile range of their respective distributions. This ensures that any remaining differences in their evolution are not driven by variations in $v_\mathrm{max}$ or $\sigma_V$.

Within these selections, we identify galaxies residing in the highest and lowest 10\% of the smoothed galaxy overdensity distribution, and further split each sample into central and satellite galaxies. This results in four subsamples for LTGs and ETGs, respectively. For each subsample, we trace the merger-tree evolution of 100 galaxies and analyse their redshift evolution.

\subsection{Mass assembly as a function of environment}

Figure \ref{fig:merger} shows the median redshift evolution of stellar, gas, and dark matter mass for LTGs (left panels) and ETGs (right panels). Shaded regions indicate the standard deviation across the 100 galaxies in each subsample. Solid and dotted lines correspond to central and satellite galaxies, respectively, while darker (lighter) colours denote galaxies residing in high- (low-) density environments.

As expected, ETGs are statistically more massive and more gas-poor than LTGs, reflecting earlier quenching of star formation. Beyond these global trends, a clear environmental dependence is present for all mass components. For both ETGs and LTGs, central galaxies in high-density environments form earlier and begin accreting mass at higher redshift than their counterparts in underdense regions. As a result, their gas reservoirs are consumed earlier and star formation peaks at higher redshift.

Satellite galaxies typically form earlier than centrals, consistent with hierarchical structure formation. However, their subsequent evolution differs markedly due to environmental processes. In dense regions, satellites experience accelerated quenching, driven by a combination of tidal stripping, ram-pressure stripping, and repeated interactions within massive halos \citep{Wetzel_2010, Watson_2012, Chaves_2016}. While the detailed realism of these processes in hydrodynamical simulations remains uncertain, the qualitative trends are robust across our samples.

For LTGs, the evolution of satellite galaxies is relatively insensitive to large-scale environment once $v_\mathrm{max}$ is fixed. Although modest differences in gas and dark matter content are visible, the final stellar mass shows little dependence on overdensity. This suggests that, for LTGs, environmental effects primarily modulate the timing of star formation rather than its integrated outcome.

In contrast, ETG satellites exhibit a much stronger environmental dependence. Satellites in dense regions form stars earlier and more rapidly, reaching larger stellar masses at intermediate redshifts. However, they subsequently lose significant amounts of gas and dark matter, and in some cases stellar mass, due to stripping within their host haloes. As a result, at fixed velocity dispersion, satellite ETGs in high-density environments end up with lower $M_\mathrm{gas}$ and $M_\mathrm{DM}$, but higher $M_*$, compared to those in low-density environments.

\subsection{Implications for TF and FP residuals}

At fixed $v_\mathrm{max}$ or $\sigma_V$, the assembly histories of galaxies—and in particular the evolution of their stellar mass—are therefore correlated with the surrounding large-scale environment. Since stellar mass directly influences a galaxy’s luminosity and size, these trends naturally imprint correlated residuals in the TF and FP relations.

The standard TF and FP relations encode only zeroth-order information about a galaxy’s star formation history through kinematic and structural parameters. The results presented here demonstrate that variations in star formation histories driven by environment persist at fixed $v_\mathrm{max}$ or $\sigma_V$, and provide a physical explanation for the density-dependent residuals and clustering signals identified in Sections \ref{sec:pv} and \ref{mtng:sec:residual}. Incorporating additional tracers of stellar mass growth and star formation activity is therefore a natural route to reducing correlated scatter in distance–indicator relations.

\section{Conclusion}
\label{sec:conclusions}

In this work, we investigated the intrinsic scatter of the Tully–Fisher (TF) and Fundamental Plane (FP) relations and the systematic effects these residuals induce on inferred peculiar velocities and their clustering. Using early- and late-type galaxies from the MTNG hydrodynamical simulation, we constructed TF and FP relations and characterised the statistical properties of their residuals.

We found that TF and FP residuals exhibit significant correlations with the matter density field on small scales, demonstrating that the scatter is not purely stochastic. We further identified opposite systematic trends between central and satellite galaxies. For late-type galaxies, satellites tend to have underestimated brightness and peculiar velocities, while centrals show the opposite behaviour. For early-type galaxies, satellites instead exhibit overestimated sizes and velocities, whereas centrals are underestimated. These trends indicate a clear connection between residuals, environment, and galaxy type.

By analysing galaxies in fixed bins of $V_\mathrm{max}$ and $\sigma_V$, we showed that variations in stellar mass and star formation rate explain a substantial fraction of the residual scatter. However, this is insufficient to fully remove the correlation with the local environment, particularly for FP residuals. This implies that, at fixed kinematic properties, galaxy luminosities and sizes still depend on additional internal properties that correlate with environment.

To mitigate these effects, we constructed extended TF and FP relations including stellar mass, gas mass, and (s)SFR. These extended relations significantly reduce both the overall scatter and the correlated component of the residuals, especially for late-type galaxies. In contrast, calibrating separate relations for central and satellite galaxies does not consistently improve the results, indicating that the environmental dependence cannot be absorbed by a simple population split.

Tracing galaxies back through their evolutionary histories, we showed that even at fixed $V_\mathrm{max}$ or $\sigma_V$, key internal properties—most notably the star formation history—depend on local density fluctuations. The standard TF and FP structural parameters therefore capture only zeroth-order information about galaxy evolution and do not fully encode environment-dependent assembly processes.

These results have important implications for peculiar velocity studies based on TF and FP distance indicators. Correlated residuals introduce systematic effects in the inferred velocity field and its non-linear clustering, potentially biasing cosmological analyses on small scales if left unmodelled. Future applications should therefore carefully control for galaxy selection and environmental effects.

Since early- and late-type galaxies are only approximately standard objects, TF and FP calibrations may need to incorporate additional structural or evolutionary indicators to mitigate correlated residuals. If such quantities cannot be robustly measured in a distance-independent way, an alternative approach is to model the residuals phenomenologically as an effective bias expansion of the density field. Future work should also explore alternative environment definitions, redshift evolution, and observational systematics to assess the robustness of these effects in real data.

\begin{acknowledgements}
      The project leading to this publication has received funding from the Excellence Initiative of Aix-Marseille University - A*MIDEX, a French ``Investissements d'Avenir'' program (AMX-20-CE-02 - DARKUNI). REA acknowledges support from project PID2024-161003NB-I00 from the Spanish Ministry of Science and support from the European Research Executive Agency HORIZON-MSCA-2021-SE-01 Research and Innovation programme under the Marie Skłodowska-Curie grant agreement number 101086388 (LACEGAL). S.B. is supported by the UKRI Future Leaders Fellowship [grant numbers MR/V023381/1 and UKRI2044.
\end{acknowledgements}

\bibliographystyle{aa}
\bibliography{Biblio}

\appendix

\section{Measuring galaxy peculiar velocities}
\label{appendix:TF_FP}

This appendix summarises how galaxy peculiar velocities are inferred from the Tully–Fisher (TF) and Fundamental Plane (FP) relations, and how the associated intrinsic scatter propagates into distance and velocity uncertainties.

\subsection{Tully-Fisher relation}
Late-type galaxies (LTGs) are star-forming, rotationally supported systems that typically reside in low-density environments, where major mergers are relatively rare. Their dynamics are dominated by ordered rotation rather than random motions, with rotational kinetic energy exceeding the internal velocity dispersion \citep{Chisari_IAl_2015}.

LTGs exhibit a tight empirical correlation between their absolute magnitude $M$ and their maximum rotation velocity $V_{\rm max}$, known as the Tully–Fisher (TF) relation \citep{TullyFisher_1977}:

\begin{equation}
M = a \ \log V_\textrm{ max} + c,
\label{eq:TF}
\end{equation}

\noindent where $(a,c)$ are the TF coefficients. In what follows, we adopt the convention $v_{\rm max} \equiv \log V_{\rm max}$ and focus on the $r$-band absolute magnitude $M_r$ \citep{DESI_pv_survey_2023}.

Observationally, $V_{\rm max}$ is inferred from the galaxy rotation curve via Doppler shifts of emission lines, typically corrected for inclination. The TF relation is calibrated using galaxies with independently known distances (e.g. Cepheid hosts), providing a predictive model for $M_r$ as a function of $v_{\rm max}$.

The absolute magnitude is related to the observed apparent magnitude $m_r$ and luminosity distance $D_L$ as

\begin{equation}
M_r(z) = m_r - 5\log\left(\frac{D_L(z)} {1 h^{-1} \text{pc}}\right) + 5,
\label{eq:M_z}
\end{equation}

\noindent where $z$ is the spectroscopically measured redshift. Because redshift-space distortions (RSD) perturb the observed redshift, peculiar velocities bias the inferred luminosity distance.

Using Eqs. \ref{eq:rsd} and \ref{eq:M_z}, we define the {\it luminosity log-distance ratio} as

\begin{equation}
\eta_L \equiv \log \frac{D_L(z_\textrm{ obs})}{D_L(z_\textrm{ cos})} =0.2 \left(M_r(z_\textrm{ cos}) - M_r(z_\textrm{ obs}) \right).
\label{eq:eta_L}
\end{equation}

Once calibrated, the TF relation provides an estimate of $M_r(z_{\rm cos})$ from $v_{\rm max}$, allowing $\eta_L$ to be inferred from measurements of $m_r$, $v_{\rm max}$, and $z_{\rm obs}$.

\subsection{Fundamental plane relation}

Early-type galaxies (ETGs) are quenched, dispersion-supported systems that typically reside in denser environments and have experienced significant merger activity. Their internal dynamics are dominated by random motions rather than rotation.

ETGs follow the Fundamental Plane (FP) relation, an empirical correlation between their physical effective radius $R$, central velocity dispersion $\sigma_V$, and mean surface brightness $I$ within $R$ \citep{FundamentalPlane_1987}:

\begin{equation}
\log R = a \ \log \sigma_{V} + b \ \log I  + c,
\label{eq:FP}
\end{equation}

\noindent where $(a,b,c)$ are the FP coefficients. The surface brightness is defined as

\begin{equation}
\begin{aligned}
I &= \frac{L}{\pi R^2}\\
&=(1+z_\textrm{ obs})^4\frac{1}{\pi\theta_e^2}10^{10-0.4m} \left( \frac{648}{\pi}\right)^2
\end{aligned}
\label{eq:surf_bright}
\end{equation}

\noindent with $L$ the luminosity in $L_\odot$, $m$ the apparent magnitude, and $\theta_e$ the angular effective radius in arcseconds.

We adopt the standard notation $r \equiv \log R$, $s \equiv \log \sigma_V$, and $i \equiv \log I$. The FP relation is calibrated using galaxies with independently known physical sizes, allowing $r(z_{\rm cos})$ to be predicted from measurements of $s$ and $i$.

The physical radius is related to the angular size and angular diameter distance as

\begin{equation}
r(z) = \log \theta_e + \log \left(\frac{D_A(z)}{1 h^{-1} \text{Mpc}} \right) + \log \frac{\pi}{648},
\label{eq:r_z}
\end{equation}

\noindent where $D_A(z)$ is the angular diameter distance. Since $z_{\rm obs}$ is affected by peculiar velocity, we define the {\it angular log-distance ratio} as

\begin{equation}
\eta_A \equiv \log \frac{D_A(z_\textrm{ obs})}{D_A(z_\textrm{ cos})} =r(z_\textrm{ obs}) - r(z_\textrm{ cos}) .
\label{eq:eta_A}
\end{equation}

Thus, the FP relation provides an estimate of $\eta_A$ from measurements of $s$, $i$, $\theta_e$, and $z_{\rm obs}$.

\subsection{From log-distance ratio to peculiar velocity}

Given a measured log-distance ratio, the line-of-sight peculiar velocity $u_r$ can be inferred. Expanding the comoving distance to first order in $z_p = u_r/c$, we obtain

\begin{equation}
    D(z_\textrm{ cos}) = D(z_\textrm{ obs}) + \frac{c}{H(z_\textrm{ obs})} \left( z_\textrm{ cos} - z_\textrm{ obs}\right),
\end{equation}

\noindent leading to the distance ratio:

\begin{equation}
    \frac{D(z_\textrm{ cos})}{D(z_\textrm{ obs})} = 1 - \frac{(1+ z_\textrm{ obs})}{D(z_\textrm{ obs})H(z_\textrm{ obs})} u_r.
\end{equation}

The corresponding log-distance ratio is

\begin{equation}
    \begin{aligned}
        \eta &\equiv - \log \frac{D(z_\textrm{ cos})}{D(z_\textrm{ obs})}\\
        &= - \log\left[1 - \frac{(1+ z_\textrm{ obs})}{D(z_\textrm{ obs})H(z_\textrm{ obs})} u_r \right]\\
        & = \alpha(z_\textrm{ obs}) u_r,
    \end{aligned}
\label{eq:eta_to_v}
\end{equation}

\noindent where the last equality holds at low redshift, with 

\begin{equation}
    \alpha(z_\textrm{ obs}) \equiv \frac{1}{\ln 10}  \frac{(1+ z_\textrm{ obs})}{D(z_\textrm{ obs})H(z_\textrm{ obs})}.
\label{eq:alpha_pv_factor}
\end{equation}

Analogous expressions can be derived for luminosity and angular distances:

\begin{equation}
    \begin{aligned}
        \frac{D_L(z_\textrm{ cos})}{D_L(z_\textrm{ obs})} &= \frac{D(z_\textrm{ cos})}{D(z_\textrm{ obs})} \frac{1}{1+z_p}, \\ \\
        \frac{D_A(z_\textrm{ cos})}{D_A(z_\textrm{ obs})} &= \frac{D(z_\textrm{ cos})}{D(z_\textrm{ obs})}\left(1+z_p\right),
    \end{aligned}
\end{equation}

\indent with

\begin{equation}
    \begin{aligned}
        \eta_L & = \alpha_L(z_\textrm{ obs}) u_r, \\ 
        \eta_A & = \alpha_A(z_\textrm{ obs}) u_r,
    \end{aligned}
\label{eq:eta_TF_FP}
\end{equation}

\noindent with 

\begin{equation}
    \begin{aligned}
        \alpha_L(z_\textrm{ obs}) \equiv \frac{1}{ c \ln 10}  \left[ \frac{(1+ z_\textrm{ obs}) c}{D(z_\textrm{ obs})H(z_\textrm{ obs})} + 1 \right], \\ 
        \alpha_A(z_\textrm{ obs}) \equiv \frac{1}{ c \ln 10}  \left[ \frac{(1+ z_\textrm{ obs}) c}{D(z_\textrm{ obs})H(z_\textrm{ obs})} - 1 \right]. 
    \end{aligned}
\end{equation}

In the regime $\frac{(1+z_{\rm obs})c}{D(z_{\rm obs})H(z_{\rm obs})} \gg 1$, the three estimators satisfy $\eta \simeq \eta_L \simeq \eta_A$.

Both TF and FP relations exhibit intrinsic scatter, which propagates into distance and velocity uncertainties. Observational TF catalogues report an average distance scatter of $\sim22\%$ \citep{hong_2mtf_2019,tully_cosmicflows-4_2023}, while FP catalogues report $\sim23\%$ \citep{saulder_calibrating_2013,howlett_sloan_2022}. Since $\alpha(z)$ decreases with redshift, a fixed scatter in $\eta$ translates into increasingly large peculiar velocity uncertainties. Consequently, TF and FP peculiar velocity measurements are typically restricted to low redshift, $z \lesssim 0.1$.

\section{Correlation function estimator}
\label{appendix:estimator}
We give a complete derivation for the state of the art Landy-Szalay estimator \citep{landy-szalay} for the galaxy correlation function, and use the same methodology to derive estimators for the velocity auto and cross correlation functions. 

The correlation function of two random fields $\delta_{\rm a}$, $\delta_{\rm b}$ is defined as
\begin{equation}
\xi_{\rm ab}(\textbf{x}_1-\textbf{x}_2) \equiv \langle \delta_{\rm a}(\textbf{x}_1)\delta_{\rm b}(\textbf{x}_2) \rangle,
\label{eq:TPCF}
\end{equation}
For the clustering of galaxies, we want to estimate the over density of galaxies $\delta_g(\textbf{x})$. 
The galaxy spatial distribution is obtained by running a discrete sum $n_g(\mathbf{x}) = \sum_i^{\rm N_g} \delta^{\rm D}(\mathbf{x} - \mathbf{x_i}) \ w_{g,i} $, with $\delta^{\rm D}$ the Dirac delta function $N_g$ the number of observed galaxies, and $w_{g,i}$ the weight associated with a given galaxy $i$ to correct for systematic effects. For a complete sample without any observational effects the weights are set to $1$.
In order to derive an estimator for the galaxy correlation function, we can also write the observed galaxy density as  
\begin{equation}
n_g(\mathbf{x}) = \Bar{n}_g \mathcal{W}_g (\mathbf{x}) \left( 1 + \delta_{\rm g}(\mathbf{x}) \right) \frac{1}{w_g(\mathbf{x})},
\end{equation}
with $\Bar{n}_g$ the mean density of galaxies and $\mathcal{W}_g$ the window function describing the survey footprint.
A random catalogue $n_s(\mathbf{x})$ following the spatial distribution of galaxies is commonly used to describe the product $\Bar{n}_g\mathcal{W}_g(\mathbf{x})$. A number of random samples $N_s$ a few ten times larger than $N_g$ is usually enough to properly sample the footprint.
The random catalogue can also have associated weights $w_s(\mathbf{x})$ to reproduce observational systematics.
Rewriting the previous relation we define the galaxy overdensity field as
\begin{equation}
\delta_{\rm g}(\mathbf{x}) = \frac{w_g(\mathbf{x}) n_g(\mathbf{x}) - w_s(\mathbf{x}) n_s(\mathbf{x})}{w_s(\mathbf{x}) n_s(\mathbf{x})}.
\label{eq:delta_g}
\end{equation}
Following Eq \ref{eq:TPCF}, we square the field and assume ergodicity so the ensemble averaging $\langle . \rangle $ is equivalent to a spatial averaging.
Doing so we can derive the well known Landy-Szalay estimator \citep{landy-szalay} for the galaxy TPCF 

\begin{equation}
\xi_{gg}(\textbf{r}) = \frac{DD(\textbf{r}) -DS(\textbf{r}) - SD(\textbf{r}) }{SS(\textbf{r})} + 1,
\label{eq:LS}
\end{equation}
where $\textbf{r} = (\textbf{x}_1 - \textbf{x}_2)$ and $DD$, $RR$, $DR$ and $RD$ are the normalised auto and cross pair counts of the data and random catalogues given by
\begin{equation}
\begin{aligned}
&DD(\textbf{r}) = \langle n_g(\textbf{x}_1)n_g(\textbf{x}_2) \rangle =  \frac{1}{N_{\rm  gg}} \sum_i^{\rm N_g} \sum_{j \neq i}^{\rm N_g} \delta^{\rm D}(\mathbf{r} - |\mathbf{x}_{g,i} - \mathbf{x}_{g,j}|) \ w_{g,i} w_{g,j},\\
&SS(\textbf{r}) =  \langle n_s(\textbf{x}_1)n_s(\textbf{x}_2) \rangle =  \frac{1}{N_{\rm  ss}} \sum_i^{\rm N_s} \sum_{j \neq i}^{\rm N_s} \delta^{\rm D}(\mathbf{r} - |\mathbf{x}_{s,i} - \mathbf{x}_{s,j}|) \ w_{s,i} w_{s,j},\\
&DS(\textbf{r}) = \langle n_g(\textbf{x}_1)n_s(\textbf{x}_2) \rangle =  \frac{1}{N_{\rm  gs}} \sum_i^{\rm N_g} \sum_{j }^{\rm N_s} \delta^{\rm D}(\mathbf{r} - |\mathbf{x}_{g,i} - \mathbf{x}_{s,j}|) \ w_{g,i} w_{s,j},\\
&SD(\textbf{r}) = \langle n_s(\textbf{x}_1)n_g(\textbf{x}_2) \rangle =  \frac{1}{N_{\rm  sg}} \sum_i^{\rm N_s} \sum_{j }^{\rm N_g} \delta^{\rm D}(\mathbf{r} - |\mathbf{x}_{s,i} - \mathbf{x}_{g,j}|) \ w_{s,i} w_{g,j}.
\end{aligned}
\label{eq:pair_counts}
\end{equation}
The auto and cross normalisation factors for two catalogues $a$ and $b$ are defined as
\begin{equation}
\begin{aligned}
&N_{\rm aa} =  \left (\sum_i^{\rm N_a} w_{a,i}\right )^2 - \sum_i^{\rm N_a} w_{a,i}^2,\\
&N_{\rm ab} =  \left (\sum_i^{\rm N_a} w_{a,i}\right ) \left (\sum_j^{\rm N_b} w_{b,j}\right ).
\end{aligned}
\label{eq:pair_counts_norm}
\end{equation}

Following the same methodology we can derive Landy-Szalay-like estimators for the radial velocity auto-correlation and the galaxy-velocity cross correlation. 
As in large-scales surveys only the radial component of the galaxy velocity field can be measured, in the following any use of "velocity" implies "radial velocity".

As the peculiar velocities are measured only where galaxies are observed, the field sampled is the galaxy momentum field $p_g(\textbf{x})$. Its spatial distribution can be written as a discrete sum $p_g(\mathbf{x}) = \sum_i^{\rm N_v} \delta^{\rm D}(\mathbf{x} - \mathbf{x}_i) \ u_{g,i} \ w_{v,i} $, with $ N_v$ the number of peculiar velocity measurements, $u_{g,i}$ and $w_{v,i}$ the velocity and weight respectively evaluated at position $\textbf{x}_i$.
The galaxy momentum can also be expressed as
\begin{equation}
p_g(\mathbf{x}) = \Bar{n}_v \mathcal{W}_v (\mathbf{x})  \ u_g(\mathbf{x}) \ \left( 1 + \delta_{\rm g}(\mathbf{x}) \right) \  \frac{1}{w_v(\mathbf{x})},
\end{equation}
with $\Bar{n}_v$ the mean density of velocity measurements and $\mathcal{W}_v$ the window function describing the peculiar velocity survey footprint. Again, we can use a catalogue $n_r(\textbf{x})$ of $N_r$ random samples to describe $\Bar{n}_v \mathcal{W}_v(\textbf{x})$. 
We call the weights associated with the random catalogue $w_r(\textbf{x})$.
We rewrite the previous relation to introduce the galaxy weighted velocity field as
\begin{equation}
v(\textbf{x}) = u_g(\mathbf{x}) \ \left( 1 + \delta_{\rm g}(\mathbf{x}) \right) = \frac{w_v(\mathbf{x}) p_g(\mathbf{x})}{w_r(\mathbf{x}) n_r(\mathbf{x})}.
\label{eq:v_g}
\end{equation}
Once again, following Eq \ref{eq:TPCF} we can define the auto and cross correlation function by respectively squaring Eq \ref{eq:v_g} and multiplying it by Eq \ref{eq:delta_g} :
\begin{equation}
\xi_{vv}(\textbf{r}) = \frac{VV(\textbf{r}) }{RR(\textbf{r})},
\label{eq:LS_vv}
\end{equation}
\begin{equation}
\xi_{vg}(\textbf{r}) = \frac{VD(\textbf{r}) - VS(\textbf{r}) }{RS(\textbf{r})},
\label{eq:LS_vg}
\end{equation}
following the same convention for the pair-counts and the normalisation. Note that $\xi_{gv}$ encapsulates as much information as $\xi_{vg}$ and therefore is not used here.
In the following we refer to the "galaxy weighted velocity field"  as the "velocity field" since they are equivalent in linear theory. However the reader should keep in mind that according to our definition, the velocity correlation function written as 

$$\xi_{vv}(\textbf{r}) = \langle u_g(\mathbf{x}_1) \ \left( 1 + \delta_{\rm g}(\mathbf{x}_1) \right) u_g(\mathbf{x}_2) \ \left( 1 + \delta_{\rm g}(\mathbf{x}_2) \right) \rangle$$
should encapsulate on non-linear scales the information contained in $\xi_{gg}$, $\xi_{uu}$,$\xi_{ug}$ and higher order statistics. 

\section{Detection in SDSS peculiar velocity catalogue}
\label{appendix:SDSS}

We use the publicly available SDSS peculiar velocity catalogue \footnote{Available on \url{https://zenodo.org/records/6640513}} described in \cite{howlett_sloan_2022}.
This is the largest catalogue of distances and peculiar velocities produced to date. It is composed of 34059 ETG derived from FP measurements, extending up to redshift $z=0.1$.
We measure the cross correlation function $\xi^{vg}$ using the provided random catalogue and weights.
We use the available set of 2048 mocks reproducing the clustering,
selection function,
and expected  errors in log distance measurements to compute the corresponding covariance matrix.

\begin{figure}
  \centering
\includegraphics[width=1.\columnwidth]{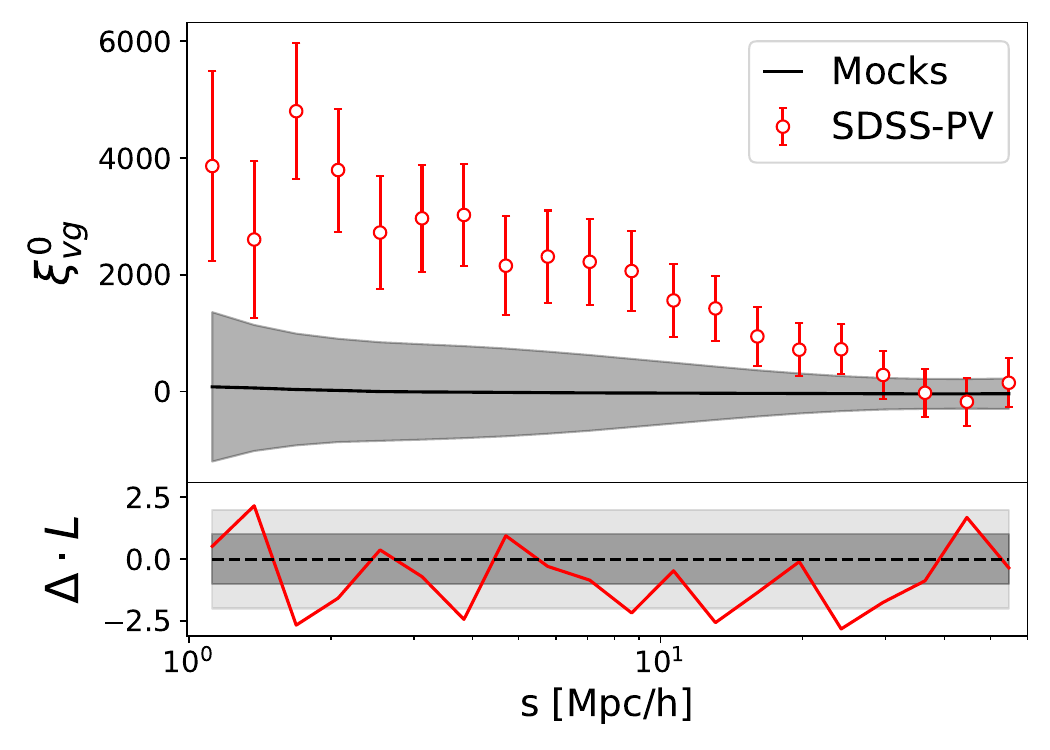} 
\caption{Correlated residual signal in the cross correlation function of the SDSS peculiar velocity catalogue. 
The black solid line indicates the expected clustering for the average of the mocks using the true velocities (consistent with zero as expected). The grey shaded area represents the corresponding cosmic variance measured for the clustering of the true velocities.
The bottom panel shows the diagonalised residuals $\Delta \xi \cdot L^{-1} $ with $L^{-1}$ the Cholesky decomposition of the inverse of the total covariance.
The grey shaded area correspond to one and two sigma deviations.}
\label{fig:sdss_data}
\end{figure}

Figure \ref{fig:sdss_data} presents the monopole of the measured clustering using the data points (in red),
with the errorbars corresponding to the covariance matrix. 
The black solid line and shaded area indicate the expected clustering averaged over the 2048 realisations using the true velocities of the mocks. 
The bottom panel shows the diagonalised residuals $\Delta \xi \cdot L^{-1} $ with $L^{-1}$ the Cholesky decomposition of the inverse of the total covariance (allowing to consider the correlations between scales in the residuals). Here $\Delta = \xi^0_{vg}$ as the expected signal is 0 for noise-free peculiar velocities.
We retrieve a non-vanishing signal, with scale dependence and amplitude quite consistent with the measurement performed using our MTNG ETG sample.
Compared to a null signal, we find a chi squared of $\chi^2/\textrm{dof} = 51.1/20$, corresponding to a p-value of $1.6\times 10^{-4}$ indicating a $3.6\sigma$ detection of correlated residuals in the SDSS peculiar velocity catalogue on the scales considered.
We leave analogous study for LTG for when the upcoming DESI peculiar velocity sample will be release \citep{saulderTargetSelectionDESI2023}. 

%
%



\end{document}